\documentclass[sigchi]{acmart}
\AtBeginDocument{%
  \providecommand\BibTeX{{%
    \normalfont B\kern-0.5em{\scshape i\kern-0.25em b}\kern-0.8em\TeX}}}

\copyrightyear{2023} 
\acmYear{2023} 
\setcopyright{rightsretained} 
\acmConference[UIST '23]{The 36th Annual ACM Symposium on User Interface Software and Technology}{October 29-November 1, 2023}{San Francisco, CA, USA}
\acmBooktitle{The 36th Annual ACM Symposium on User Interface Software and Technology (UIST '23), October 29-November 1, 2023, San Francisco, CA, USA}
\acmDOI{10.1145/3586183.3606824}
\acmISBN{979-8-4007-0132-0/23/10}

\begin{document}

\title{Never-ending Learning of User Interfaces}

\author{Jason Wu}
\email{jsonwu@cmu.edu}
\authornote{Work done while Jason Wu was an intern at Apple}
\affiliation{%
  \institution{HCI Institute, Carnegie Mellon University}
  \city{Pittsburgh}
  \state{PA}
  \country{USA}
}
\author{Rebecca Krosnick}
\email{rkros@umich.edu}
\authornote{Work done while Rebecca Krosnick was an intern at Apple}
\affiliation{%
  \institution{Computer Science and Engineering, University of Michigan}
  \city{Ann Arbor}
  \state{Michigan}
  \country{USA}
}
\author{Eldon Schoop}
\email{eldon@apple.com}
\affiliation{%
  \institution{Apple}
  \country{USA}
}
\author{Amanda Swearngin}
\email{aswearngin@apple.com}
\affiliation{%
  \institution{Apple}
  \country{USA}
}
\author{Jeffrey P. Bigham}
\email{jbigham@apple.com}
\affiliation{%
  \institution{Apple}
  \country{USA}
}
\author{Jeffrey Nichols}
\email{jwnichols@apple.com}
\affiliation{%
  \institution{Apple}
  \country{USA}
}

\renewcommand{\shortauthors}{Wu, et al.}

\begin{abstract}
    Machine learning models have been trained to predict semantic information about user interfaces (UIs) to make apps more accessible, easier to test, and to automate. Currently, most models rely on datasets of static screenshots that are labeled by human annotators, a process that is costly and surprisingly error-prone for certain tasks. For example, workers labeling whether a UI element is “tappable” from a screenshot must guess using visual signifiers, and do not have the benefit of tapping on the UI element in the running app and observing the effects. In this paper, we present the Never-ending UI Learner, an app crawler that automatically installs real apps from a mobile app store and crawls them to infer semantic properties of UIs by interacting with UI elements, discovering new and challenging training examples to learn from, and continually updating machine learning models designed to predict these semantics. The Never-ending UI Learner so far has crawled for more than 5,000 device-hours, performing over half a million actions on 6,000 apps to train three computer vision models for i) tappability prediction, ii) draggability prediction, and iii) screen similarity.
\end{abstract}


\maketitle

\section{Introduction}
Machine Learning (ML) has played an increasingly important role in the domain of mobile User Interfaces (UIs). Recent techniques have used Deep Neural Networks (DNNs) to bridge critical usability gaps and enable new types of evaluations, such as providing missing accessibility metadata to UIs~\cite{WuScreenParsing}, giving designers feedback to make UI features more discoverable~\cite{tapAmanda, SchoopTappability}, and predicting user engagement with animations~\cite{WuanimationsEngagement}.
The enabling research artifacts behind these interactions are large datasets of mobile UI screenshots annotated by human annotators~\cite{rico, frontmatterKuznetsov}. These datasets provide an invaluable volume of data for training DNNs, but they only capture a fixed snapshot of the views of mobile applications and are extremely costly to collect and update. In addition, relying on annotators to estimate certain properties of UI elements from static visual signifiers is known to be error-prone~\cite{SchoopTappability}.
Inspired by the Never Ending Learning paradigm~\cite{mitchellNELL}, we propose an automated method for collecting UI element annotations by \emph{interacting with applications directly} with an automated crawler that continuously improves its own performance and can refresh ML models for other downstream tasks over time.

We built the \textit{Never-ending UI Learner}, an app crawler that formulates UI semantic learning as an active process that uses real interactions on real devices to explore UIs and discover properties which are used to continually train machine learning models.
More specifically, our crawler automatically installs real apps from mobile app stores and crawls them to discover new, challenging training examples to learn from (e.g., those that result in low model confidence). During crawling, the Never-ending UI Learner records temporal context (i.e., taking screenshots before, during, and after interactions) that is used by heuristic functions to generate more accurate labels than are possible from human-annotated single screenshots. The resulting data is used to train models that predict the tappability and draggability of UI elements and determine the similarity of encountered screens. Although the process can start with a model trained from human-labeled data, the end-to-end process does not require any additional human-labeled examples.

In contrast to existing data pipelines for data-driven UI modeling \cite{rico,ZhangScreenRecognition,frontmatterKuznetsov}, our never-ending UI learning paradigm allows data collection, annotation, and model training to be performed without any human supervision and can be run indefinitely. Of course, in this paper the learning is not truly never-ending. Here we present experiments that analyze the performance characteristics of our learner over 5,000 device-hours, in which it performed more than half a million actions on 6,000 apps.
The resulting dataset is an order of magnitude larger than existing human-annotated UI datasets \cite{ZhangScreenRecognition,rico} and allowed us to analyze the performance of UI semantic models when trained with increasing amounts of recently collected examples. Ultimately, we believe this model can be used in a true ``never-ending'' style, continually crawling the app ecosystem, collecting data from literally all available apps, and experiencing new UI styles and trends as new or updated apps are released.

The specific contributions of our paper are as follows:

\begin{enumerate}
    \item \textbf{The Never-ending UI Learner}, is a system that operationalizes our approach for automatically learning from UIs through never-ending interaction.
    \item \textbf{Three applications which demonstrate use of the Never-ending UI Learner}. We use our crawler to train three types of models of UI semantics that are difficult to learn through existing methods: i) tappability, ii) draggability, and iii) screen similarity.
\end{enumerate}

\section{Related Work}
Our work in never-ending learning of UIs aims to supplement UI modeling datasets used to model UIs and user interaction through continual learning. To situate our work, we review related literature in the i) UI modeling datasets, ii) computational models of interaction, and iii) approaches for continual machine learning.
\subsection{Datasets for Modeling User Interfaces}
Several datasets
have been collected for the purposes of analyzing and modeling mobile UIs. The Rico dataset is a large dataset of 72,000 mobile UIs and associated metadata including view hierarchies, screenshots, and user interactions, collected from 9,700 publicly available Android apps~\cite{rico}. The FrontMatter dataset uses static analysis techniques to predict the purpose of UI elements by determining which system APIs are invoked~\cite{frontmatterKuznetsov}. Large datasets like these have enabled ML-based methods which can perform various tasks involving mobile UIs, including providing accessibility annotations~\cite{WuScreenParsing, li2020widget}, giving design feedback~\cite{swireHuang, tapAmanda, SchoopTappability, yuan2020modeling}, suggesting common interaction flows~\cite{zhou2021large}, summarizing screens~\cite{wang2021screen2words}, automating interaction with UIs~\cite{li2020mapping, sereshkeh2020vasta, arsan2021app}, and creating rich embeddings of UI image and text data for other downstream tasks~\cite{screen2vec, BaiUIBert, heActionBert}.
Almost all currently available datasets are manually created in some aspect -- through manual user interactions with UIs, and/or human annotations.
The WebUI dataset \cite{WuWebUI} used screenshots and automatically extracted metadata from web pages to train visual UI models; however, web data was generally noisy and their models needed additional fine-tuning on smaller human-annotated datasets to perform well.
Our Never-ending UI Learner produces annotations through the automated crawling of mobile applications. These annotations continually update and refresh the crawler's models, improving its performance, and resulting in a continually updated dataset that can be used to train other models. An important advantage of this approach is that, unlike using mobile UI data collected during a specific time period, data produced by our crawler is always current, and can support updating models to keep up with evolving UI design trends in mobile applications.

\subsection{Computational Modeling of Interaction}
\textcolor{black}{An important application of large annotated UI datasets is supervised training of machine learnings that predict UI semantics. For example, in this paper we focus on the problems of element tappability \cite{SchoopTappability,tapAmanda} and screen ``fingerprinting" \cite{feizScreenSimilarity}.}
More recently, Reinforcement Learning (RL) has been applied to model user interactions with both physical and digital interfaces. Oulasvirta et al. proposed a general framework based on RL of how users incorporate cognitive facilities, their experiences, and their environment in understanding and interacting with computers~\cite{oulasvirtaRationality}. 
Under this context, an important part of knowing how to interact with an interface is by understanding its affordances.
Affordances are the functional properties of an object (e.g., UI) that suggest how it should be used \cite{NormanPOET}, and designer commentary suggests that design patterns can make affordance discovery more difficult.
Liao et al. used a virtual robot agent equipped with sensors to simulate and learn how humans may discover affordances in physical interfaces (e.g., buttons and sliders)~\cite{liaoRediscoveringAffordance}.
Our work aims to achieve similar goals of learning the affordances (e.g., tappability) and capabilities of interfaces.
While our work does not directly model the interactions of users through RL techniques, we aim to achieve similar goals of learning of the affordances and capabilities of interfaces through interacting with and inspecting live mobile apps. By applying interaction learning to a mobile app automated crawler, we can scale our experiments to a much larger scale, learning from millions of interactions with UIs.

\textcolor{black}{\subsection{Continual Machine Learning}}
A unique aspect of our work is the intention to continually learn about UIs over time through sustained, potentially endless interaction. 
Our work is related to active learning (specifically online active learning), which is a field of ML that seeks to improve models using only a limited number of human-labeled examples~\cite{DeepLearningBookGoodfellow}. These approaches often identify and prioritize difficult or representative examples to produce the best possible model from a small dataset.
Our work is most related to Never-Ending Learning, which is an ML paradigm for creating systems that continually learn from acquired experience rather than a single dataset. It was first applied to web-based knowledge using the NELL system~\cite{mitchellNELL}.
The system has been running for prolonged periods of time (years) and has accumulated over 50 million beliefs (i.e., hypothesized knowledge snippets), which is possible only by processing large amounts of data that are prohibitively expensive to annotate.
This learning approach introduces unique challenges, such as the need to learn from new data while retaining previously acquired knowledge.
There are several techniques in the literature that can be applied to retain previous knowledge that involved i) regularization \cite{LiLearningWithoutForgetting,kirkpatrick2017overcoming}, ii) rehearsal-based approaches \cite{rebuffi2017icarl}, and iii) techniques that address task-recency bias \cite{castro2018end}.
From a practical standpoint, implementation also necessitates maintaining large ever-growing datasets collected over time, which could either be addressed through a robust crawling infrastructure or using dataset distillation methods that keep the most relevant samples \cite{wang2018dataset, nguyen2020dataset,nguyen2021dataset}.
In this work, we apply the never-ending learning paradigm to benefit automated UI understanding systems by training models ``from scratch'' and fine-tuning existing models to improve performance.

\section{Never-ending UI Learner}
To operationalize our approach, we built the Never-ending UI Learner, a system that automatically downloads and crawls publicly available apps using remotely operated devices. Our current implementation and infrastructure is based on iOS. We use stock factory reset devices that are logged in to testing accounts that are not associated with any real user data to avoid privacy concerns.

Note that unlike some crawlers that interact with apps using an OS-provided programmatic interface such as the accessibility API, our crawler interacts with the device through the VNC remote desktop protocol, from which it receives regular updates to the screen and processes them visually and can send raw input events to the device to create tap, swipe and keyboard actions. Using VNC, the Never-ending UI Learner is able to reliably interact with more apps, learn based on the same facilities that a human would and generalize to other platforms. In this section, we describe the crawler’s architecture and behavior that enable it to perform never-ending learning.

\subsection{Architecture Overview}
  \begin{figure}
    \centering
    \includegraphics[width=18pc]{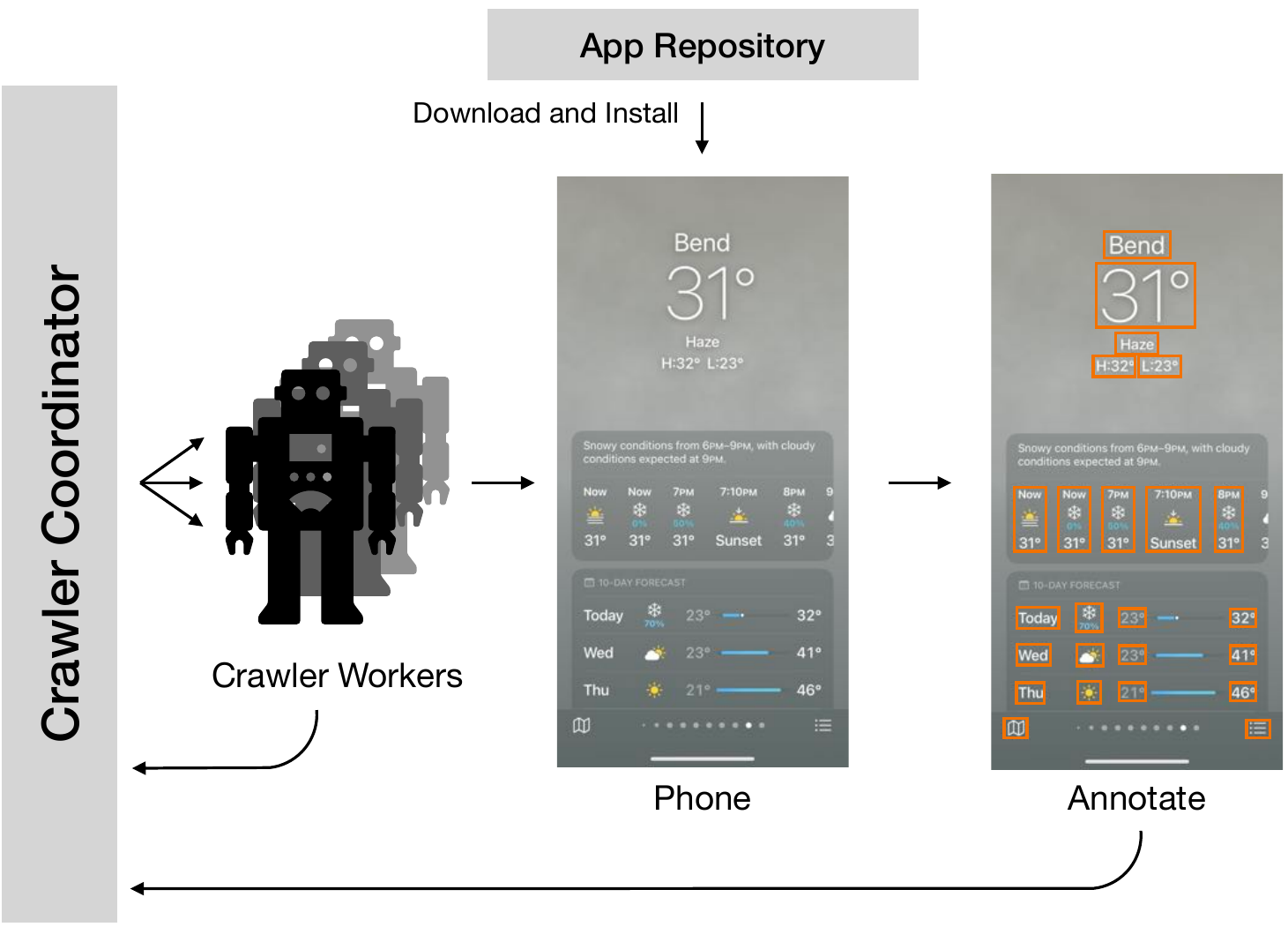}
    \caption{
    Architecture of our Never-ending UI Learner. The Never-ending UI Learner is a parallelizable mobile app crawler which consists of a coordinator-worker architecture. The crawler coordinator distributes crawls to workers and maintains the dataset. Each crawler worker is connected to a programatically controlled mobile device which collects data and runs data post-processing.}
    \label{fig:crawler_architecture}
\end{figure}
Our crawling architecture is shown in Figure 1. We implemented a distributed crawling architecture which consists of i) a central coordination server and ii) a large pool of workers to parallelize the crawling process.

\subsubsection{Coordinator Server}
The crawler coordination server maintains a list of app IDs to crawl which are sent to workers. The central server keeps track of successful and unsuccessful crawls, and it automatically retries failed app crawls. App crawlers differ from web crawlers in that they focus only on the app they are asked to crawl, although limited cross-app interaction sometimes does occur (e.g., clicking on a link or permission request dialog). When all app IDs are exhausted, our crawler can schedule itself to be run again after a fixed time period (e.g., weekly). The list of app IDs can be modified between crawls to add new apps or reflect changes in app availability. While the majority of the app IDs remain the same, the apps may change their appearance and behavior due to dynamically updated content and new versions of the software. Re-crawling the same apps regularly can enable our model to adapt to design changes over time.

\subsubsection{Crawler Worker}
Crawler workers are processes that interface with remotely controlled mobile phones and process the collected data. Each crawler worker downloads and installs a target app whose ID is provided by the central server to the mobile phone and then runs a program that crawls the app. Screenshots are collected during interactions and when the crawler believes it has arrived at a new screen. \textcolor{black}{The program can use three methods to explore the app (random selection or based on model confidence), and }as a part of this paper, we run experiments to determine the best crawling strategy for each of our never-ending learning use-cases. We set a time-limit (5 minutes) on the maximum duration of a crawl for a single app. Afterwards, the worker processes the collected data (e.g., screenshots and interactions) with models and heuristics to generate labels from the observations. Both raw data and processed output is uploaded to a coordinator server. \textcolor{black}{In our experiments, the number of crawler workers varied to due to availability from the device pool which we used, which was shared with other users. Generally, there were around 40-100 crawler workers.}

\subsection{Machine Learning Components}
Our crawler contains a screen-level and element-level model that allow it to understand the content on UIs it encounters. We run these models every time a screenshot is captured to augment it with useful semantics. Furthermore, the three UI semantic models that we trained in using the crawler, are designed as extensions of these base models, improving overall efficiency.

\subsubsection{Screen Understanding}
To keep track of its crawling progress in the app, our crawler uses a model to generate semantic representations of screens. We used a model introduced by previous work \cite{feizScreenSimilarity} that predicts whether two screenshots belong to the same UI by encoding each as an embedding vector, which the authors shared with us. Because significant variation can be introduced by changes in state, such as a news app that displays new content periodically, the model is designed to learn the underlying structure of UIs.
We made minor modifications to the previous work in order to develop a model that could run under our hardware constraints.
Instead of their recommended \textit{screen transformer} model architecture, we use their CNN-based model architecture, which is more efficient to run despite somewhat lower performance \cite{feizScreenSimilarity}. For further optimization, we use an \textcolor{black}{EfficientNet-B0} \cite{TanEfficientNet} model architecture as the backbone instead of the original ResNet-18 \cite{HeResNets}, which has more parameters. \textcolor{black}{As in the original paper, the output of the last layer of the CNN network is used as a screen embedding.} During training, we applied a data augmentation approach \cite{thakur2020augmented} to increase performance. We followed all other aspects of the original model training and our final CNN-based model achieves a F1-score of 0.636.

\subsubsection{Element Understanding}
To generate element semantics, we used an object detection model architecture that is similar to CenterNet~\cite{zhouCenterNet2019}. At a high level, the detection model slides a window (via convolutions) over the image and featurizes image sub-regions using a backbone network (MobileNet-v1~\cite{howard2017mobilenets}), resulting in embeddings for each region. These embeddings are fed into a classification head which produces per-class confidences, and regions with high confidences are returned as detections. The model was trained on the \textsc{AMP} dataset \cite{ZhangScreenRecognition}, which consists of ~77,000 app screens collected and annotated by annotators from ~4,000 iPhone apps. In addition to the standard element type classification head, which was trained with the rest of the object detection model, we added heads for tappability prediction and draggability prediction. The additional heads are trained independently from the rest of the model by first freezing the backbone and training the heads on embeddings corresponding to detected elements.

\section{Applying Never-ending Learning} %
In this section, we describe the application of our never-ending learning framework to three UI semantic models: i) tappability prediction (element semantic), ii) draggability prediction (container semantic), and iii) screen similarity (screen semantic).
The tappability and draggability models were trained completely from crawler-generated data, while the crawler fine-tuned its existing screen similarity model that was originally trained using human-annotated data.
For each UI semantic, we developed an interaction-based heuristic used by our crawler to automatically generate new training examples for our models. Next, we designed and trained models to predict each of these semantics from a screenshot. Finally, to contextualize these models in the context of never-ending learning, we analyzed their performance over time.

\textbf{Experimental Setup. }
We conducted experiments on a list of 6,461 free iOS apps.
For the purposes of evaluation, all model training and experiments were performed with randomized training (80\%), validation (10\%) , and testing (10\%) splits.
We randomly partitioned our list of app IDs, which ensured that all UI screens from an app were contained in the same split. We use the term \textit{crawl epoch} to refer to one complete pass through the list of apps. Note that unlike an \textit{epoch} through a training dataset, the actual contents of a \textit{crawl epoch} might change from time to time, due to the dynamic nature of apps.

Our experiments analyzed two aspects of the crawler's performance: i) crawling strategy and ii) performance over time.
We ran three variations of the crawler, which had different crawling strategies: i) randomly selecting elements on each screen (Random), ii) selecting elements that result in low prediction confidence from the current models (Uncertainty Sampled), and iii) a hybrid that for each crawl epoch alternates between Random and Uncertainty Sampled strategies, \textcolor{black}{inspired by similar approaches in optimization \cite{smith2017cyclical}}.
To evaluate the performance over time, we ran each crawling strategy for five crawl epochs.
Note that the first crawl epoch for all strategies uses Random to train an initial confidence-prediction model.  In the Hybrid strategy, because alternation happens at the epoch level, the second epoch is crawled using the Uncertainty Sampled strategy and thus through two epochs the inputs and results are identical for both the Uncertainty Sampled and Hybrid strategies.  The three strategies fully diverge starting from the third epoch. \textcolor{black}{Across all experiments, we collected over half a million screenshots, although the same UI screen may have been visited multiple times. The number of screenshots in our dataset is considerably larger than previous work \cite{bunian2021vins,frontmatterKuznetsov,WuWebUI,rico}.}

The crawler's models were trained and evaluated after each crawl epoch. \textcolor{black}{After each crawl epoch, model training is resumed with the updated data from the latest crawl and the model weights are optimized} for 100,000 optimization steps (with early stopping).
In order to maintain a constant validation set across a varying number of epochs, we only use the evaluation data split from the first epoch for calculating performance metrics.
Finally, for models that were trained completely on crawler data (tappability and draggability), we performed additional sub-epoch evaluations during the first crawl to analyze learning speed.

\textcolor{black}{
While the dataset is not released at the time of publication due to internal regulations, we are investigating processes to make it available to the broader community.
To replicate our work, it is possible to use tools and models built for comparable platforms (e.g., Android). Open-source crawlers \cite{li2017droidbot,toyama2021androidenv} can be integrated with available screen similarity \cite{WuWebUI} and element detection models \cite{WuWebUI,bunian2021vins,xie2020uied}.}

\subsection{Tappability}
Tapping is the most common interaction on mobile devices, yet it is often difficult to automatically determine if an element is tappable or not due to missing metadata and ambiguous visual cues. \textcolor{black}{For example, a text button that doesn't have sufficient contrast or missing borders would likely appear untappable to users, and many games are missing accessibility traits that prevent screen reader users from using them.} Accurate inference of tappability could aid designers in finding ambiguous visual elements and be useful for generating metadata for repairing inaccessible apps. Previous work has used human-annotated UI screenshots to train machine learning models of tappability. However, this process is surprisingly error-prone~\cite{liDenoise22, enrico, tapAmanda, SchoopTappability} due to ambiguous visual cues, which suggests that human-annotated screenshots are an unreliable source of ground-truth for training tappability models. In contrast, our crawler can use additional context from the entire interaction, such as before and after screenshots instead of a single before screenshot, to determine if tapping resulted in an effect.  Effects could either be state changes, like flipping a toggle, or a transition to a new screen. We developed a heuristic for inferring tappability from our crawler’s recorded interactions and found that it had high agreement with human-annotated videos. We used heuristic-labeled data to train an efficient tappability ``head" model purely from crawler-annotated data. After five crawl epochs, the best-performing tappability model reached an F1 score of 0.860.

\subsubsection{Tappability Heuristic}
  \begin{figure}
    \centering
    \includegraphics[width=\linewidth]{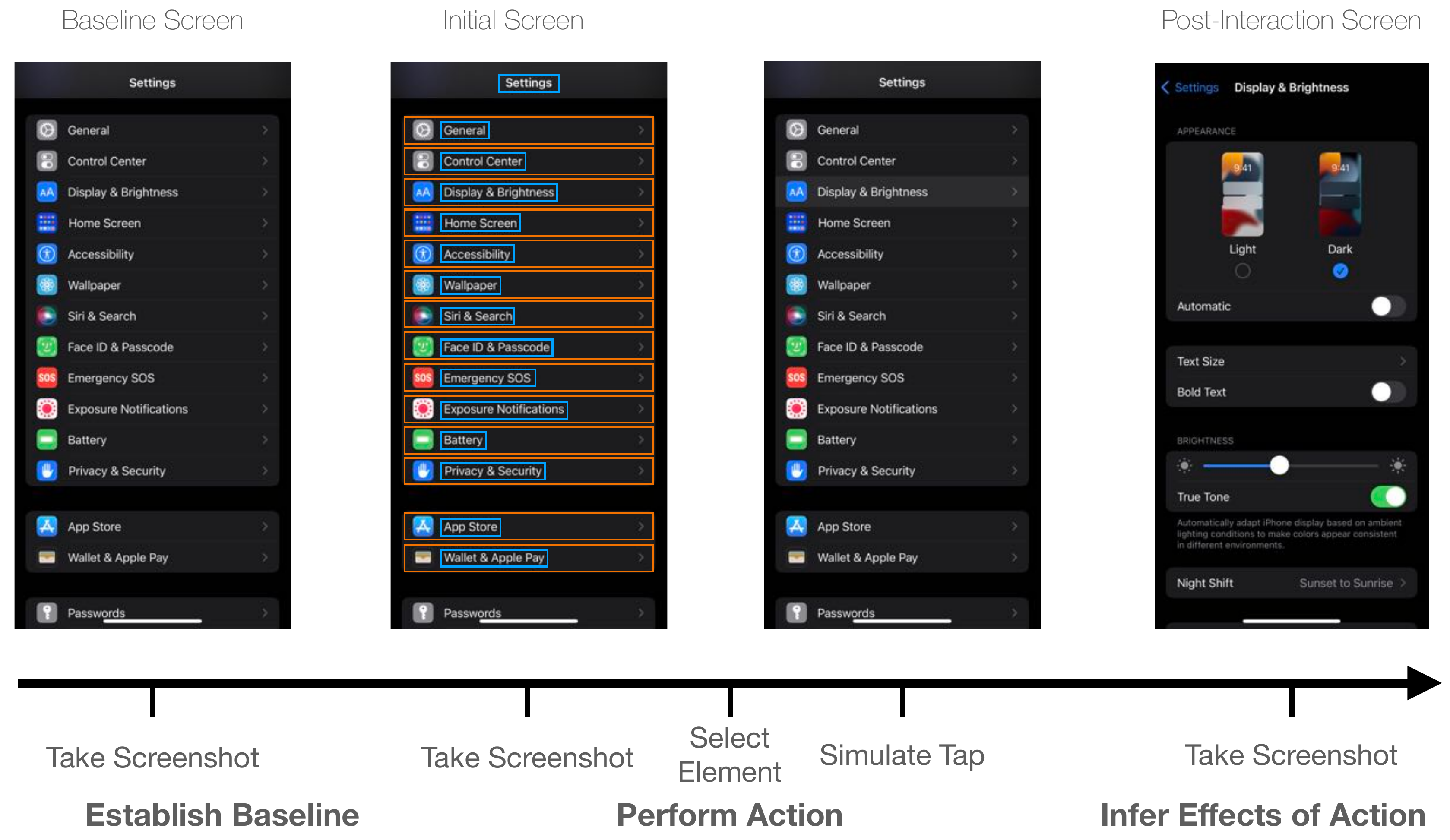}
    \caption{
    This figure visualizes the steps to our tappability heuristic. When the crawler arrives at a new screen, it takes two screenshots separated by 5 seconds as a baseline of visual change. Then, a detected UI element is chosen and sent a tap. After waiting for the screen to settle, a post-interaction screenshot is used to infer the effects of the action.}
    \label{fig:tappability_heuristic}
\end{figure}

We developed a heuristic to infer the tappability of an element based screenshots of the UI taken before, during, and after a tap interaction. A tap may result in several different scenarios, which are captured by our heuristic. First, we use a screen similarity model to compare screenshots taken before and after the tap to determine if the tap led the crawler to a new screen. If a screen change was not detected, the tap could have also changed the screen state. We compute a pixel-based difference of the “before” and “after” screenshots to identify possible visual indications of local or global changes, such as tapping a checkbox or refreshing screen content respectively. Finally, to reduce false positives, the heuristic also uses multiple screenshots captured before the tap to identify dynamic areas of the screen (e.g., videos) whose visual changes are not related to the tap.

To validate the accuracy of our heuristics, we compared its results against human-labeled interaction videos. We used our crawler to save short screen recordings of tap interactions that were collected during crawls. Each example video was approximately 10 seconds long and included the tap location overlaid on the video and temporal context before and after the tap interaction, such as including transition and loading animations.

We randomly sampled a balanced subset of 1000 video clips from our crawls and asked human annotators if each video clip contained a tap interaction. \textcolor{black}{Annotators were recruited, trained, and paid by a separate team at our organization (all with appropriate legal/ethical approval). Annotators were employees paid who are paid a competitive hourly salary for their location.} We used standard classification metrics to evaluate the accuracy of our heuristics, using the human-annotated labels as ground truth.
The tappability heuristic had an overall accuracy of 0.934, and had a similar number of false positives (38 instances) and false negatives (28 instances).

\subsubsection{Model Implementation}
  \begin{figure}
    \centering
    \includegraphics[width=\linewidth]{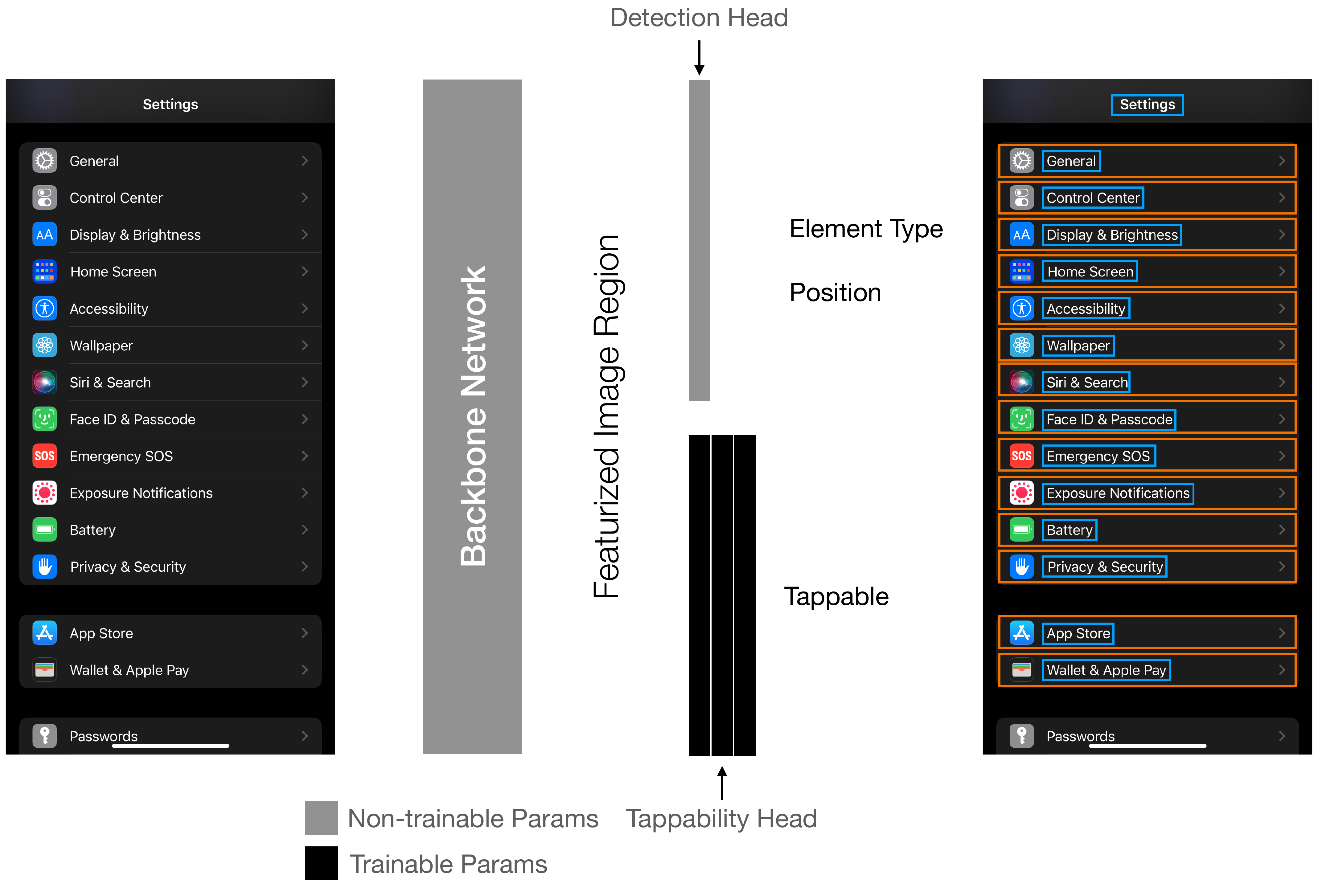}
    \caption{
    Architecture of our tappability model. The tappability model is designed as a ``head", which is a sub-network of the UI element detection model. The element detector featurizes image regions in an input screenshot using a sliding window, which results in a featurized image embedding for each detected object. The main branch of the network (top) feeds in the embedding to determine the region's element type and position. We feed in the same element embedding into a separate feedforward network (bottom) to predict the probability that it is tappable.}
    \label{fig:tappability_model}
\end{figure}
To predict tappability, we designed a model architecture that operates as a ``head'' of our existing element detection model (Figure \ref{fig:tappability_model}).
Heads are small sub-networks or set of layers usually located close to the output layer of neural network architectures and generate predictions from featurized representations of the main input produced by a “backbone” network. Since element detection is closely related to tappability, we hypothesized that the previously learned representations are likely to contain relevant information and greatly accelerate tappability learning. Our head model is a simple three-layer feed-forward model with an input size of 128, a hidden size of 64 that we chose through manually tuning, and an output size of 1 that gives tappability confidence. To train it, we first froze the weights of the element detector's backbone network and randomly initialized the parameters of our feed-forward network. While freezing most of the model reduces its capacity, it also results in a significant reduction in training time, since there are fewer parameters to optimize.
Then we trained the model to predict the tappability of an element from a screenshot of the UI before the tap, and we used the labels generated by our tappability heuristic as ground truth.\textbf{}

\subsubsection{Performance Evaluation}

  \begin{figure}
    \centering
    \includegraphics[width=\linewidth]{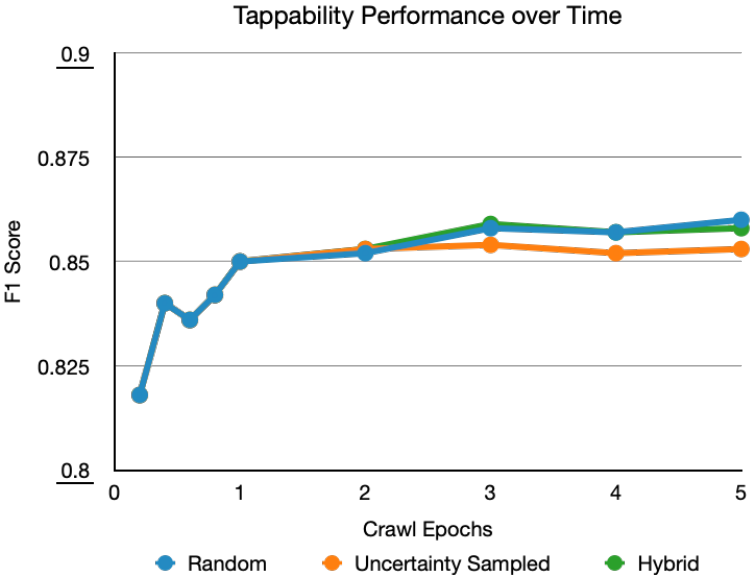}
    \caption{
    Performance of tappability over time. The model performance increases most rapidly during the first crawl epoch and the rate of improvement plateaus afterward. After the final epoch, the random crawler achieves the highest F1 score of 0.860, and the uncertainty sampled crawler has the lowest F1 score of 0.853.}
    \label{fig:tappability_performance}
\end{figure}

The results of our experiments are shown in Figure \ref{fig:tappability_performance}.
While all crawling strategies are successful in improving on the initial model from the first epoch, the Random crawler has the best final performance.
In our experiments, the Random crawler reaches the best final F1 score of 0.860 while the Uncertainty Sampled crawler reaches the lowest final F1 score of 0.853.
While it is not possible to make a direct comparison with previous work \cite{SchoopTappability,tapAmanda} because their experiments were run on different datasets, it seems that our tappability model is able to reach similar levels of performance in terms of F1 score after its first epoch.

We also conducted a comparison between the quality of our automatically collected tappability dataset and human-annotated ones, we used the labels provided by the \textsc{AMP} dataset \cite{ZhangScreenRecognition}.
First, we trained our classification head model architecture on \textsc{AMP}, which led to similar performance (F1=0.81) to the originally reported numbers (also F1=0.81), which used a tree-based model architecture.
However, when we used the model trained on human-annotated data to predict the tappability of elements in our crawled dataset, we observed significantly degraded performance (F1=0.60), suggesting that the human-annotated and crawler-generated labels disagree with each other.
We consider the heuristic-annotated data to be higher quality since its performance was validated by annotators with access to a video clip of the entire tapping interaction, and previous work \cite{SchoopTappability} has shown predicting element tappability from a single screenshot leads to high variance among raters.
\subsection{Draggability}
Dragging is a common interaction in mobile apps that involves touching an element on the screen with one's finger and moving the finger along the screen's surface before finally releasing it. This interaction is used to manipulate controls, such as sliders and page controls, and is necessary for accessing off-screen content via scrolling. While these examples reflect different types of input, we collectively refer to all these actions as ``draggability,'' since they involve similar physical movement. Unlike tappable elements, draggable elements often have fewer visual signifiers and are more difficult to automatically detect. To the best of our knowledge, there aren’t any datasets available with draggability labels, and we believe that, similar to tappability, it would be difficult for human labelers to reliably identify draggable elements from screenshots. To improve screen reader support for inaccessible apps with these affordances, we used our crawler to automatically find and label examples through automated interaction.
 We developed a heuristic to infer draggability from screenshots of drag interactions.
 Using data labeled by this approach, we trained a draggability model that reached an F1 score of 0.794 after five crawl epochs.

\subsubsection{Draggability Heuristic}
  \begin{figure}
    \centering 
   \includegraphics[width=20pc]{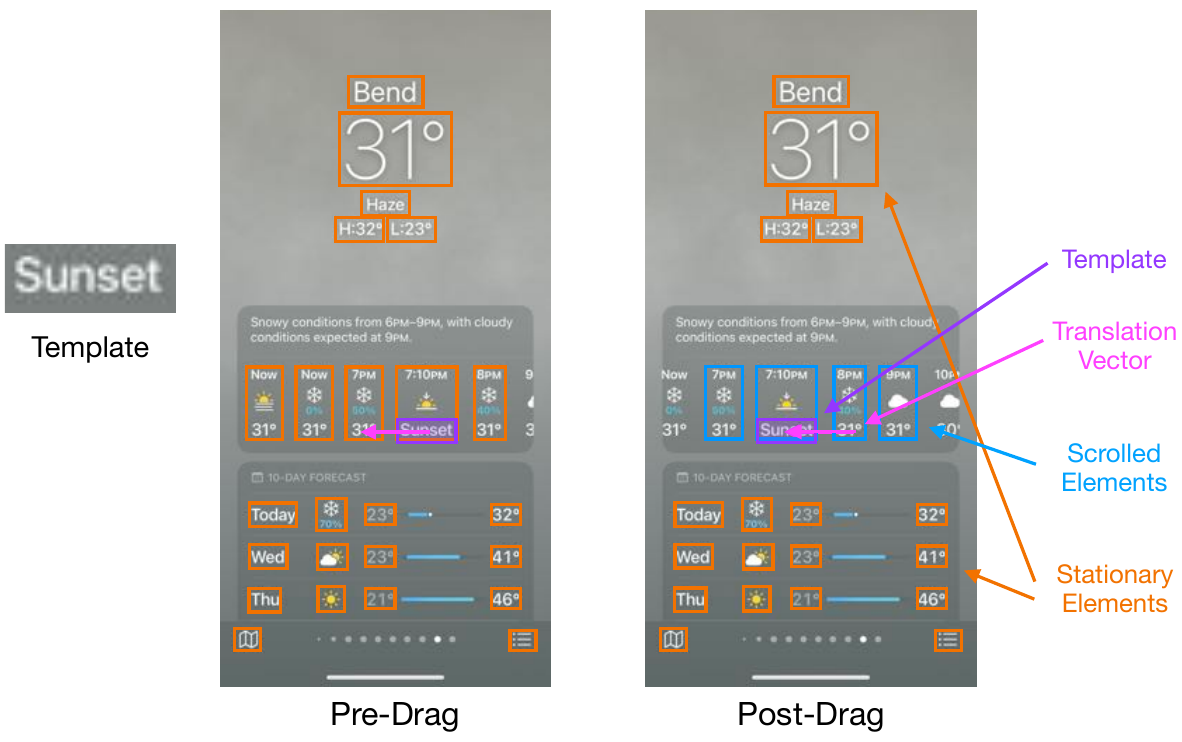}
    \caption{This figure illustrates the draggability heuristic. The heuristic uses a pre-drag image (center) which was taken before the interaction, and a post-drag image (right) which is taken near the end of the drag interaction, before the ``finger" leaves the screen. A template image is created from the dragged element (left). The heuristic finds the location of the template in the post-drag image to infer draggability.}
    \label{fig:draggability_heuristic}
\end{figure}
To detect if a UI element is draggable, our crawler captures screenshots while attempting to hold and drag elements. Our crawler first identifies likely candidates, then emulates drag actions to the left (e.g., finger goes to the left) and upward directions to detect horizontal and vertical dragging, respectively. These directions were chosen because they correspond to the initial position of lists in left-to-right reading directions, and we execute drag actions from the center of the element to either its left or top boundary.
The crawler captures one screenshot before the drag begins and one screenshot at the end of the drag but before its ``finger" leaves the screen.

The high-level idea of the heuristic is to detect which UI elements, if any, ``follow the finger" in the direction of the drag. We first use the smallest UI element containing the dragged pixel on the pre-drag image to create an image patch. This image patch is template-matched with the post-drag image using the normed correlation coefficient method on grayscaled and edge-detected images. The vector corresponding to the template displacement is filtered by cosine angle and magnitude.
Next, the patches inside bounding boxes between the pre-drag and post-drag screens are compared to identify whether other elements which scrolled during the drag action. If the contents of a bounding box in the pre-drag image match the contents of a bounding box \emph{translated by the template translation vector} in the post-drag image, then it is likely a UI element which has been scrolled together with the original UI element. We use the normed correlation coefficient method to measure similarity between these image patches, on grayscaled and edge-detected images. If no scrolled elements are identified, the original UI element is also marked as not draggable to filter out false positives.

We conducted an evaluation of our heuristic on 1000 samples, which were generated by running the heuristic on screens collected from a randomized crawl, then selecting 500 screens where the heuristic was triggered and 500 where it was not.
Due to a glitch, our crawler did not record the interaction videos of the draggability interaction, however we found that it was straightforward to infer draggability from the captured before/after screenshots.
For each interaction step, we presented the annotator with three images, the pre-drag image, post-drag image, and a combined image with the both pre- and post- images super-imposed, which allowed more easy visualization of movement.
The images were annotated with an arrow that indicated where the drag occurred.
Again, we used the human-provided labels as ground-truth to evaluate our heuristic's predictions.
The draggability heuristic had an overall accuracy of 0.92, and a similar number of false positives (38 instances) and false negatives (48 instances).

\subsubsection{Model Implementation}
  \begin{figure}
    \centering
    \includegraphics[width=\linewidth]{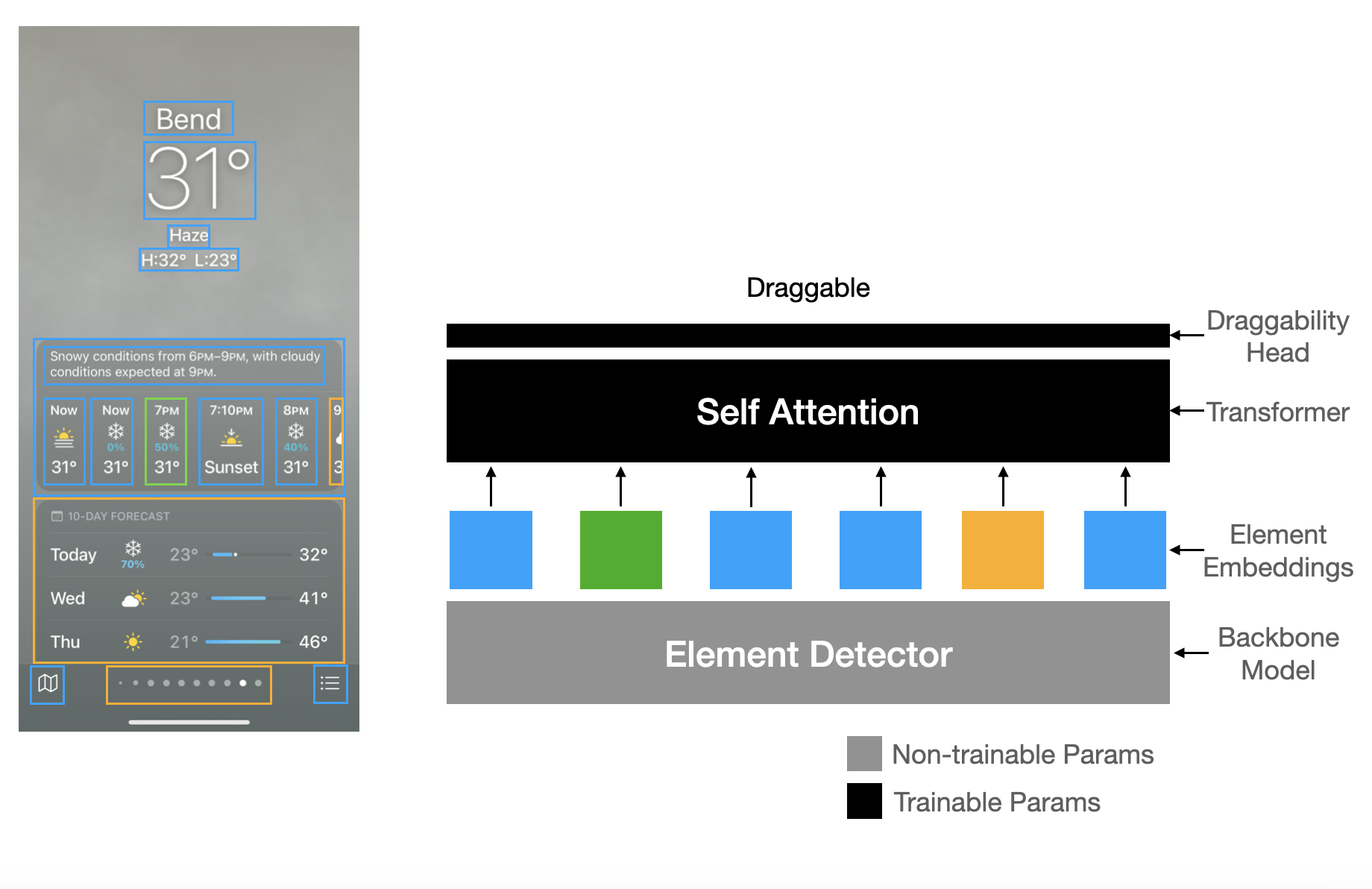}
    \caption{
    \textcolor{black}{Architecture of our draggability model. Similar to the tappability model, the draggability model uses embeddings from the element detector. To give the draggability model additional context (e.g., presence of partially occluded elements), all elements on the screen are simultaneously fed into a single-layer transformer. The resulting contextual embeddings are used to predict draggability probability.}}
    \label{fig:draggability_model}
\end{figure}

Unlike tappability, which is an element semantic, draggability is often associated with containers that contain multiple elements.
We initially tried to use the same ``head'' model architecture as our tappability model, which re-uses the element features generated by our detection model, however we found that this model achieved low performance (F1 score=0.2). Upon closer inspection of misclassified examples, we noticed that visual signifiers for an element's draggability are often non-local (i.e., occur elsewhere on the screen). For example, a picture is more likely to support swiping if page control indicator is located beneath it, and scrollability in mobile apps is often best inferred by searching for partially occluded elements at the end of the list or near the edges of the screen. Because the element detector featurizes image regions by pooling together nearby visual information, it omits many relevant cues for this task.

Based on these observations, we designed a model based on the transformer architecture, which allows it to incorporate information from the entire screen into its prediction. We first used our element detector to featurize all detected elements on the screen. The element embeddings are then fed into self-attention layers to generate a contextualized embedding. Finally, the elements' contextualized embeddings are fed into a linear classifier with a single output node to classify draggability. While training the draggability model, the element detector's weights are also frozen to improve training efficiency. For screens where the draggability heuristic wasn't triggered, loss is only computed for the directly interacted element. For screens where the draggability heuristic was triggered, loss is computed on all elements that were affected by the drag. In both cases, elements that did not move along with the finger are ignored in the loss calculation, as it isn't possible to know for certain if they are not draggable without interacting with them.

\subsubsection{Performance Evaluation}

\begin{figure}
    \centering
    \includegraphics[width=0.9\linewidth]{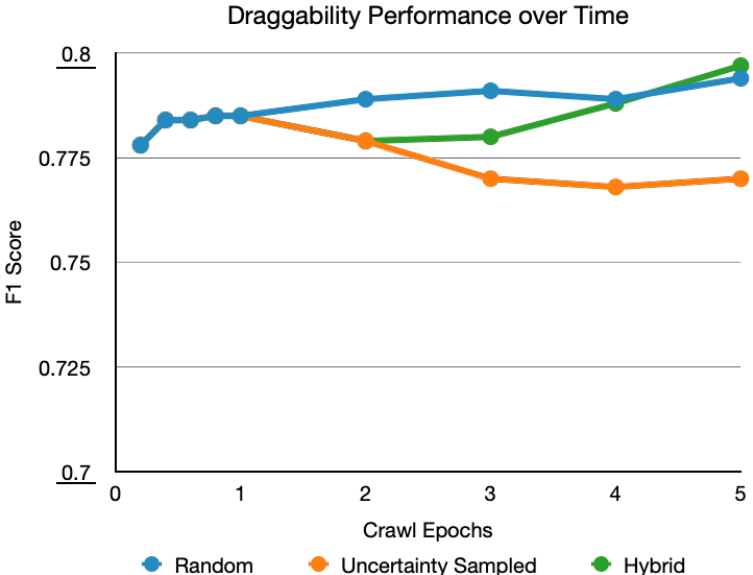}
    \caption{
    Performance of draggability over time. Similar the tappability model, the draggability model improves the most during the first crawl epoch and the rate of improvement plateaus afterward. The hybrid strategy crawler achieves the highest final F1 score of 0.797, and the uncertainty sampled crawler has the lowest F1 score of 0.770.}
    \label{fig:draggability_performance}
\end{figure}

Our evaluation of the draggability model focused on performance over time (See \autoref{fig:draggability_performance}).

The results of our experiments are shown in Figure \ref{fig:draggability_performance}.
The Hybrid crawler had the highest final performance (F1=0.794), while the Uncertainty Sampled crawler was lowest (F1=0.770).
Interestingly, the Uncertainty Sampled and Hybrid crawls both experienced a decrease in performance during the second crawl epoch.
While the Uncertainty Sampled crawler continued to decline, the Hybrid crawler alternated to its randomized crawl strategy and began to rapidly improve.
We hypothesize that the uncertainty sampling during the second epoch may have imbalanced the dataset by collecting many examples of similar elements while ignoring others, and thus negatively impacted the subsequent model.

From our experimental results and anecdotal observations, we hypothesize that draggability is harder to infer from static visual information alone due to the lack of local cues, and the best way to discover functionality that involves dragging may be learning from extended usage.
In some cases, it may be appropriate to directly apply the draggability heuristic at run-time.
In contrast to tapping, which is likely to alter the state of the UI or bring the user to a new page, we hypothesize that many dragging interactions are less likely to lead to side-effects.
Our model could be used to first identify likely candidates for interaction-based verification.

Similar to the tappability model, we also observed small gains in performance over time; however, there was less overall improvement to draggability performance.
One possible reason is that since draggability is more difficult to infer visually, the model reached its ceiling earlier.

\subsection{Screen Similarity}
We used our crawler to improve its screen understanding capabilities by using its interactions to validate and retrain the screen similarity model. A more accurate screen similarity model allows our crawler to more reliably determine which app screens it has already visited in an app, and thus increase its exploration efficiency.
Screen similarity models have also been used in other types of software engineering applications, such as processing mobile app usage videos \cite{CooperDuplicateVideo}, automated software testing \cite{LiDroidBot,LiHumanoid}, and automated storyboard generation \cite{ChenStoryboard}.
Feiz et al. note that due to their labeling technique, their dataset contains more examples of new-screen pairs than same-screen pairs. We mined additional examples of same-screen pairs from our crawler's recorded interactions to augment the original training data and fine-tuned the initial model by lowering the learning rate by a factor of 10. Compared to a baseline condition where the screen similarity model was trained using the unmodified dataset \textcolor{black}{(with the same lowered learning rate)}, we found that the augmented dataset led to consistently better performance.

\subsubsection{Data Generation}
We do not introduce a new interaction-based heuristic for collecting labels for screen similarity. Instead, we re-use the data captured from the tappability and draggability heuristics. Both heuristics take two screenshots before initiating an interaction to identify animated or dynamic regions of the screen that could cause false \textit{positives} for tappability and draggability detection. Yet these same examples can also be used to find examples of false \textit{negative} predictions from our screen similarity classifier. We assume that the pre-interaction screenshots belong to the same screen, since any visual variation between them is not caused by a user input. We make similar assumptions about data collected from the draggability heuristic, since the final screenshot is taken before the drag gesture is completed (i.e., before the finger is released from the screen) and is unlikely to result in a new screen. We use these sources to create a dataset of screenshot pairs of same-screen pairs, and we ran our existing screen similarity model to search for incorrect predictions, which can be used to re-train the model. Based on this process, we mined approximately 2000 new examples from each epoch.
\subsubsection{Model Implementation}
  \begin{figure}
    \centering
    \includegraphics[width=\linewidth]{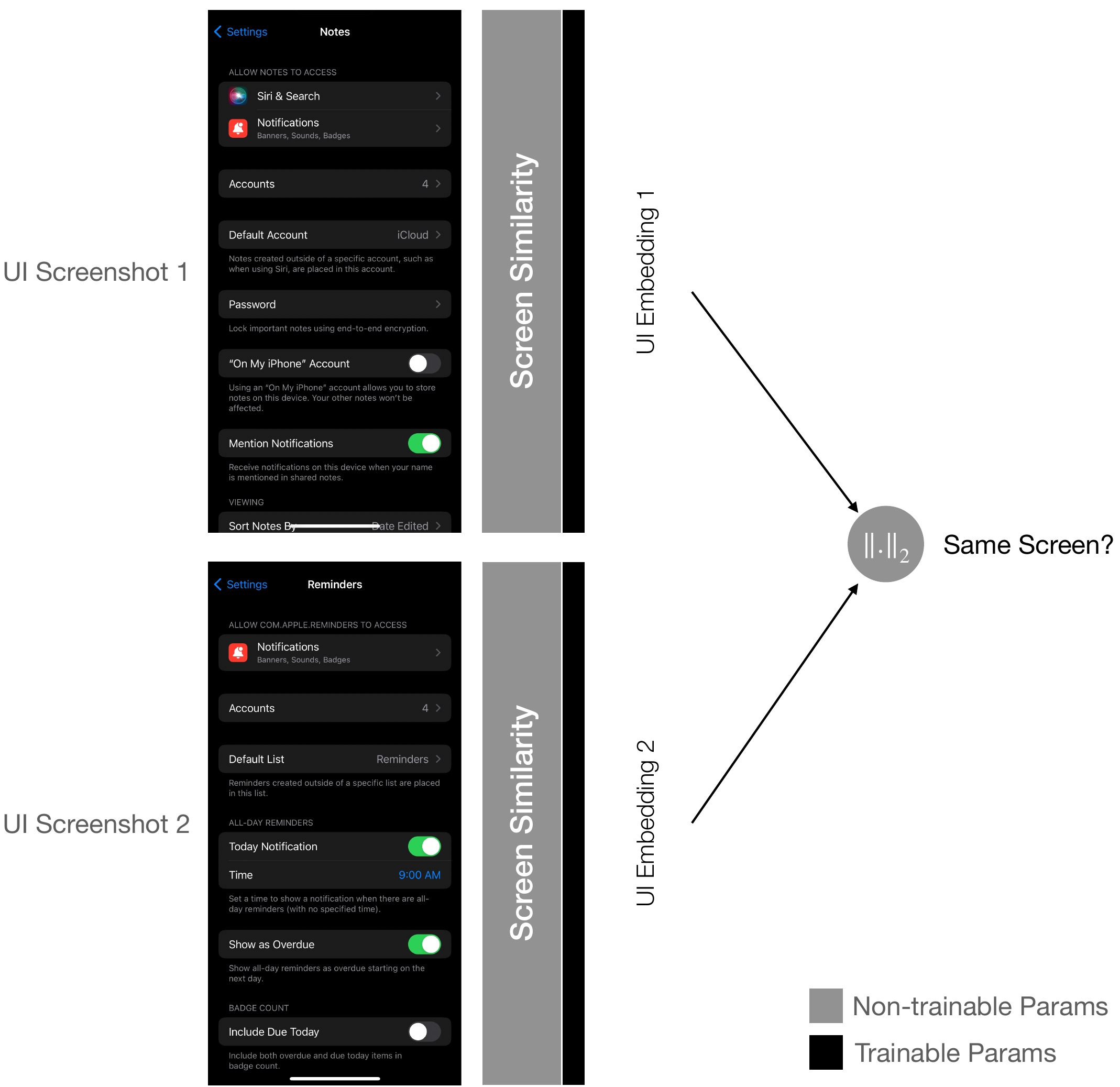}
    \caption{
    Architecture of our screen similarity model. The screen similarity determines if two input screenshots are variations of the same UI by i) featurizing each screenshot using a CNN ii) comparing their Euclidean with a threshold value.}
    \label{fig:screensim_model}
\end{figure}
The screen similarity model was initially trained on a dataset that contained both examples of positive (same-screen) and negative (different-screen) pairs, which made it possible to optimize using a contrastive margin loss \cite{HadsellInvariantMapping}.

At a high level, the model maps screenshots into an embedding space, and the loss ensures that similar screens are close together (i.e., have a distance less than a margin value) while different screens are further apart.
\begin{equation}
  \mathcal{L}_{sim} =
  \begin{cases}
    {||\Delta h||}_2 & \text{if $s_1 = s_2$} \\
    \mathrm{max}(0, m-{||\Delta h||}_2) & \text{otherwise}
  \end{cases}
\end{equation}
To fine-tune the model, we use the same training objective but decrease the learning rate to $lr=1e-5$, which is ten times lower than the original value used to train the model.
We initially tried to use the newly-mined same-screen pairs to fine-tune the model without mixing it with the original dataset. However, this was unsuccessful, since only focusing on the ``similarity'' term ($s_1 = s_2$) resulted in a failure case where the model learns to map all screenshots to the same point in embedding space, since it is only penalized if similar screens are far away but not if dissimilar screens are close together.
Thus, we directly ``mixed" in the newly mined examples with the rest of the original dataset, which consisted of ~800,000 labeled pairs.

\subsubsection{Performance Evaluation}
\begin{figure}
    \centering
    \includegraphics[width=\linewidth]{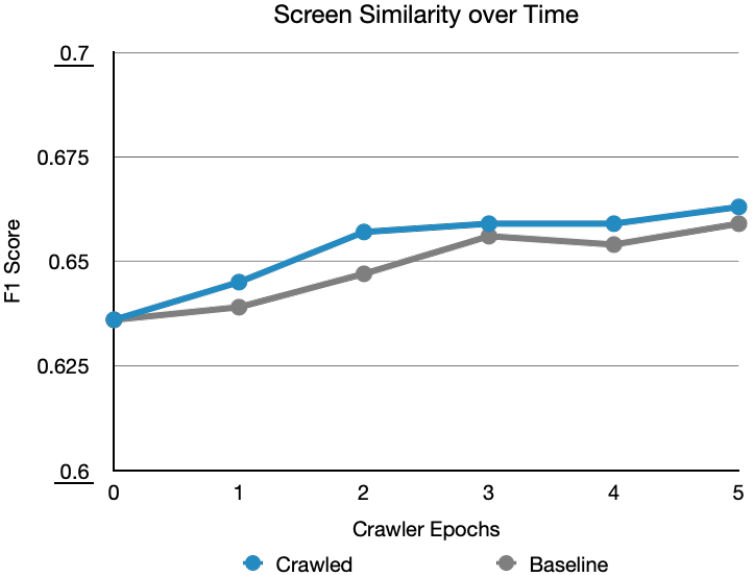}
    \caption{
    Performance of screen similarity over time. We compared i) adding training examples mined from crawls and ii) a baseline of continuing model training on its original dataset with a lower learning rate. The crawler-augmented dataset achieved a final F1 score of 0.663 while the baseline's final F1 score was 0.659.}
    \label{fig:screensim_performance}
\end{figure}

We measured performance with respect to the original dataset's evaluation split because our generated data only contains same-screen pairs, which makes it impossible to compute precision.
The results are shown in Figure \ref{fig:screensim_performance}.
Because the screen similarity model doesn't affect the crawler's selected actions (e.g., attempted taps and drags), we only evaluated our approach on data from the Random crawl.
Overall, we found that using the crawler-generated dataset to fine-tune the model led to small but consistent improvements in performance over time.
The screen similarity model improved from an initial F1 score of 0.636 to a final F1 score of 0.663.
Despite being trained on the original dataset, the baseline model also improved due to the lowered learning rate. A common practice in neural network training is to decrease the learning rate after performance plateaus, which may allow the model to continue improvement. The baseline model improved the initial model to a final F1 score of 0.659.

If the crawler were to run indefinitely, it would need some mechanism to ignore a subset of the collected data to avoid eventual data imbalance due to the collection of only same-screen pairs. Several possible techniques exist for consolidating and distilling datasets to retain the most informative samples \cite{nguyen2020dataset,nguyen2021dataset,wang2018dataset}. While we believe these methods are applicable, we leave this aspect of validation to future work.

\section{Discussion}
Our experiments revealed that visual UI models could effectively be trained and improved through automated, continual interaction.
In this section, we discuss i) the performance of our specific Never-ending UI Learner implementation, ii) other types of interaction-based learning, and iii) the benefits applying these strategies over very long or potentially indefinite period of time
\subsection{Never-ending UI Learner Performance}

In this paper, we conducted a series of experiments that evaluate the Never-ending UI Learner and and its ability to automatically learn UI semantics. Our experiments investigate two key questions: i) what is the best way for an automated crawler to learn about UIs? and ii) how long would it need to run?

\textit{Crawling Strategy. }
Our experiments focused on three crawling strategies for exploring mobile apps: i) randomized crawling, ii) uncertainty sampling, and iii) a hybrid strategy.
Overall, the random strategy consistently led to strong performance in all our experiments.
We initially hypothesized that uncertainty sampling, an active learning technique that improves sampling efficiency by prioritizing examples with low model confidence, would let the model to learn more efficiently and effectively.
However, because our crawler updated its models (which are used to compute the prediction confidences) every epoch instead of after each sample (as is often done in applications where uncertainty sampling is employed), it led to imbalanced data collection during subsequent crawls, which decreased performance.
The hybrid crawler alternated between random and uncertainty sampling strategies, which allowed it learn from low-confidence predictions while also correcting the distribution shift induced by batched uncertainty sampling. Overall, it led to similar performance to pure random crawling, although it was less consistent. In the draggability task, it initially decreased performance but experience rapid improvement afterward.
Ultimately, our experiments do not reveal a clear choice, and we believe there is room for exploring additional strategies \cite{DeepLearningBookGoodfellow} and longer-term evaluation, which we leave to future work.

\textit{Performance over Time. }
Even though our crawler is meant to run indefinitely, our experiments focused on a relatively short period of five crawl epochs.
Each crawl epoch lasted approximately half a week (clock time) when parallellized across multiple crawler workers and consisted of approximately 500 device-hours of app interaction, data post-processing, and model training. Across all experiments, the Never-ending UI Learner crawled for more than 5,000 device-hours, which was carried out over the span of approximately one month.

Our results show that this window is sufficient to learn accurate models purely from crawler-collected data (tappability and draggability) or fine-tune existing models (screen similarity).
Overall, we found that models had rapid early learning followed by slower improvement, which is consistent with empirical observations in machine learning research that suggests an exponential relationship between dataset size and model performance
\cite{SunUnreasonableEffectivenessofData}.
We believe these small improvements are valuable, since their benefit can be magnified when running over potentially very long periods of time and allow the model to be continuously updated.
We plan to continue running the crawler, which doesn't require human supervision, to observe trends over longer periods of time and maximize the potential of our automated learning approach.

\textcolor{black}{\textit{Anecdotal Observations. }}
\textcolor{black}{Based on our experimentation, we found several dimensions that affect the performance of never-ending learning systems such as ours. We offer anecdotal observations that may be useful for replication or implementing similar systems.}
\begin{itemize}
    \item \textcolor{black}{\textit{Choosing examples. } In this paper, we primarily explored two methods (random and uncertainty-based) for selecting training examples. We found that random selection is a strong baseline, and active learning approaches (e.g., uncertainty sampling) can be effective with proper hyperparameters. There is more to explore along this dimension, including the use of crawler history to reduce sample redundancy, which we observed in our crawled data.}
    \item \textcolor{black}{\textit{Retraining frequency. }Re-crawling and re-training frequency can affect the system's performance changing the makeup of the training data. The experiments in this paper were run over the span of around one month with an update iteration every 1-2 days, so the app changes that we witnessed were primarily due to changes in dynamic content. We believe that less frequent updates can be effective (e.g., monthly) which could capture more substantial changes such as app updates or newer design guidelines.}
    \item \textcolor{black}{\textit{Evaluation data. } We used a fixed evaluation split to directly compare model performance over time. Using data from the latest crawl may allow for more accurate estimation of real-world performance; however, it is less straightforward to compare models across epochs, since changes may either have been caused by model performance or changes to validation data. Finally, using a dynamic-sized validation set could be useful if the models are deployed in scenarios where performance on both old and new apps are important.}

\end{itemize}

\subsection{Learning from Interactions}
Our work contributes the idea that automated interactions can be used to generate datasets for model-based UI understanding.
Most existing datasets use human annotators and crowd workers to produce labels for mobile UI datasets, such as UI element bounding boxes, icon types, and screen similarity pairings. While human labeling has been a de facto standard for creating datasets, especially for domains where the data volume requirements for self-supervised learning are not feasible, crowd worker-generated annotations are known to be susceptible to errors and biases~\cite{ClarkHumanGold, ParmarBlameAnnotator}. Furthermore, many tasks implicitly encode a degree of subjectivity. One such example is tappability prediction, where annotators use their own judgment to decide whether a particular UI element in a screenshot is tappable. Labels for such tasks are known to be noisy in practice, and are often averaged or voted on from multiple crowd workers, further increasing the time and cost of human-produced labels~\cite{SchoopTappability}.
In contrast, automated interaction-based learning can significantly mitigate annotator biases, since labels are produced through hypothesis testing. However, there may be cases where encoding perceptual information into labels can be useful, such as giving feedback to designers on \emph{perceived} tappability. In other cases, such as generating accessibility information, labels more closely aligned with ground truth may be preferred. Understanding the trade-offs between these methods and their impact on model alignment is an opportunity for future work.

In our work, we showed that interaction-based learning can be used to model element (tappability), container (draggability), and screen-level (screen similarity) semantics in mobile UIs.
For our tested applications, we found that heuristics that operated with knowledge of the entire interaction made label generation relatively straightforward.
However, highly accurate heuristics did not always lead to highly accurate models since the model had to make the same prediction with access to less data (i.e., only visual information from a static screenshot). Some types of semantics were more conducive to visual modeling than others.
Our tappability model achieved high classification performance, with an F1 score of 0.860.
On the other hand, draggability was much harder to predict from a screenshot (F1=0.797).

A natural question to explore is: what other types of semantics can be learned through interaction?
For example, related semantics such as ``press-and-hold'' functionality can be discovered, and textboxes can be better understood by observing what kind of software keyboard (e.g., email or numeric keyboard) appears when it is tapped on.  Could this approach be extended to the problem of UI element detection more generally, which currently relies heavily on human annotation?  There are many details that would need to inferred, such as the size and shape of UI elements, and of course the element type.  Many more interactions would be needed from the crawler to determine a bounding box for a given element, and it might be difficult to infer complex element types, but a working system that could do this might be able to learn about custom controls and other non-standard elements that current models cannot deal with today.
Better automated understanding of UIs can not only benefit downstream applications directly, but also collect better data to train models.

\subsection{Benefits of Never-ending Learning}
Our crawler is meant to be run indefinitely, allowing it to accumulate examples and train over long periods of time.
In our paper, we experimented with several variables (e.g., training hyperparameters and exploration strategies), which was only feasible by focusing on a relatively short period of time for each condition (5 crawl epochs).
Even from this short time-span, we could train models for UI semantics ``from scratch'' and observed consistent improvements to performance afterwards, but we believe that our models are yet to reach their maximum potential.
In addition to its performance benefits, never-ending learning allows machines to learn from diverse sources of data.
Never-ending learning can help machines identify and learn from mistakes, especially those caused by shifts in data distribution caused by trends in app usage and design trends.

Never-ending learning also introduces new challenges, like ``catastrophic forgetting,'' the possibility of erasing previously learned information by training on new data, and difficulties associated with large, ever-growing datasets.
In this paper, we conduct a preliminary exploration of methods to address some of these challenges, such as uncertainty sampling, which can help prioritize certain types of data.
Our literature review uncovered many other possible machine learning techniques that involve training the model training process \cite{LiLearningWithoutForgetting,kirkpatrick2017overcoming,rebuffi2017icarl,castro2018end} or distilling the collected dataset relevant samples \cite{wang2018dataset, nguyen2020dataset,nguyen2021dataset}.
We expect that they will be useful for scaling and maximizing the performance of never-ending UI learning.

\section{Limitations \& Future Work}
Our current implementation of a Never-ending UI learner is limited and presents opportunities for future exploration.

First, our current crawler is implemented using a specific set of tools and infrastructure customized for our target platform (iOS). While we did not run experiments on other types of UIs (e.g., Android, web-based interfaces), we expect our results to be generalizable, since our approach does not rely on any platform-specific metadata or APIs, and previous research has shown semantic overlap between mobile and web UIs \cite{WuWebUI}. %
Our experiments primarily focused on free apps that did not require authentication (e.g., registering and making an account), which biased the set of UI screens reached by crawling.
We used manually-designed and verified heuristics for a small set of semantics for tappability and draggability. We believe that many other aspects of UIs and interaction can be formulated using similar methods.
\textcolor{black}{Another limitation of our current experiments is that we did not investigate the effect of different randomized train/test splits, which could provide additional insight into the robustness of our method. Because experiments took roughly a month to complete, the time and compute costs for repeated trials would have been prohibitively high. However, since our list of apps is sufficiently large and randomly shuffled, we do not expect large variations in performance across different randomized splits.}

Personalized interaction traces collected over long periods of usage can improve the performance of models for rarer, niche apps, although a privacy-preserving approach would be needed (e.g., on-device training). An alternative direction is to allow our crawler to automatically learn interaction sequences to discover and label new aspects of UIs, instead of executing manually-defined heuristics. We expect future versions of our crawler to incorporate techniques from related machine learning fields, such as reinforcement learning. 

Finally, our crawler could benefit improved UI understanding capabilities.
First, our crawler's primary representation of screens that it visits is a list of UI elements, which are used to navigate and discover other parts of the app.
A more effective way of representing screens could lead to more efficient crawling \cite{WuScreenParsing}. For example, since properties of list items are similar, the crawler could reduce unneeded interactions by tapping on one list item and propagating the label to others.
Icon semantics \cite{liu2018learning,chen2020unblind,chen2022towards} are also helpful for inferring the result of certain interactions. For example, tapping on a ``camera" icon may open the system camera app, which would disrupt the crawl.
Since the goal of the crawler itself is to train such UI models, we believe that integrating these additional models into the never-ending learning framework is a natural next step.

\section{Conclusion}
In this work, we presented a technique for continuous extraction and modeling of user interface semantics through interactions, which we refer to as ``never-ending learning of UIs.'' We implemented a mobile app crawler that downloads, installs, and crawls thousands of apps to observe UI semantics and affordances in real-world apps, and we use interaction-based heuristics to generate large datasets for training three types of UI understanding models i) tappability, ii) draggability, and iii) screen similarity. We found that models trained in this way can be more accurate than those trained from human-annotated screenshots and continue to improve with access to more training examples. The highly automated nature of our approach allows us to apply it indefinitely, with little to no human supervision, which can maximize their performance and utility to downstream applications.

\bibliographystyle{ACM-Reference-Format}
\bibliography{sample-base}

\appendix
\textcolor{black}{\section{Annotation Interface}}
  \begin{figure*}[!htb]
    \centering
    \includegraphics[width=0.95\textwidth]{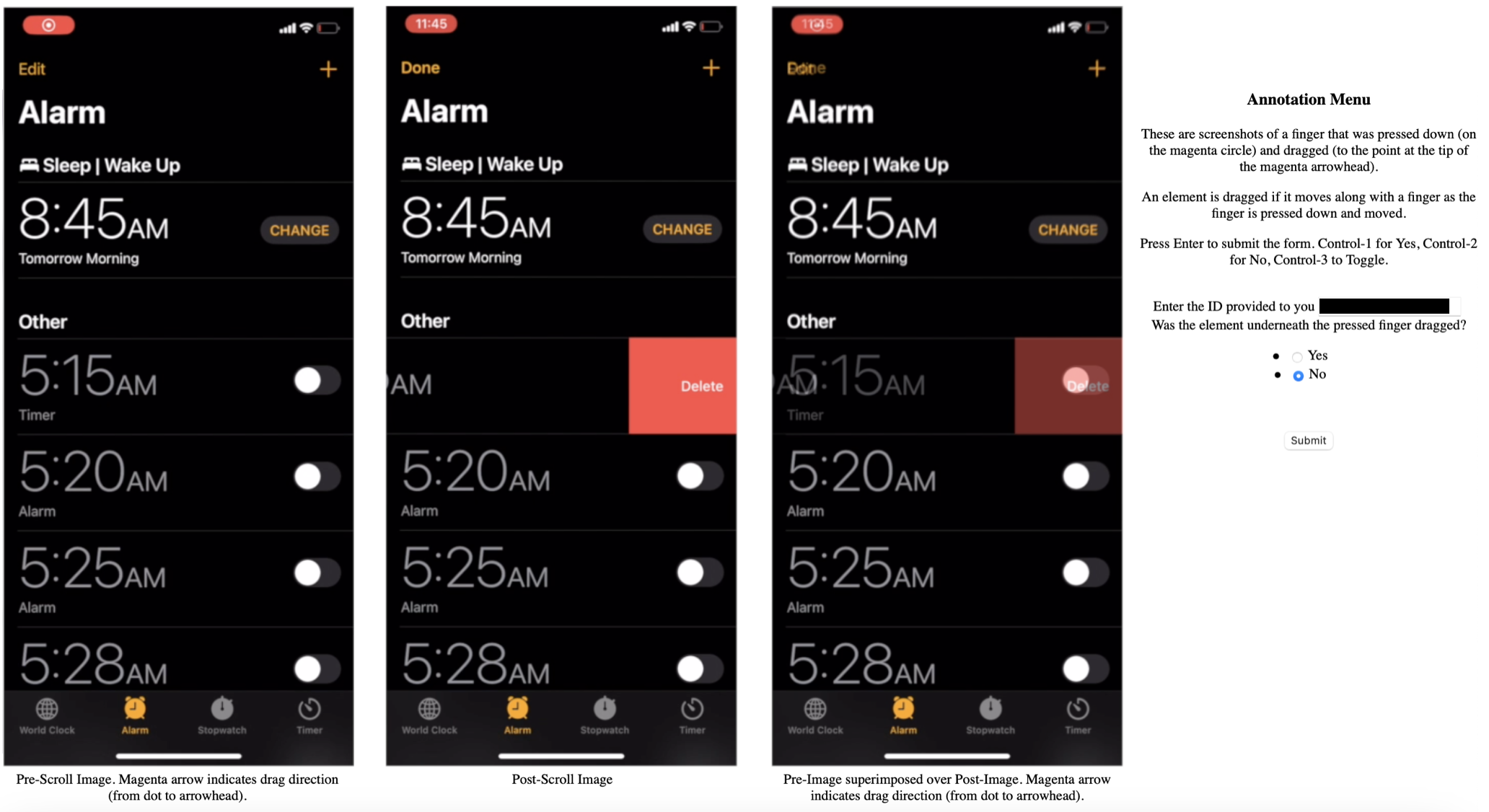}
    
    \caption{
    \textcolor{black}{Figure shows the web-based interface used for the draggability heuristic evaluation, which displays the state of the screen before the drag, during the drag, and a superimposed image. The interface used for tappability evaluation is similar, except that the annotators watched a video clip instead.}}
    \label{fig:annotation_interface}
\end{figure*}
\textcolor{black}{Annotators used a web-based annotation interface to evaluate the performance of the labeling heuristics. Figure \ref{fig:annotation_interface} shows the web-based interface used to label heuristic outputs.}

\textcolor{black}{\section{Model Hyperparameters}}
\textcolor{black}{Table \ref{tab:hyperparams} shows the hyperparameters used to train the various models in the Never-ending UI Learner.}
\begin{table*}[!htb]
\caption{\textcolor{black}{Hyperparameters of the models used in the Neverending UI Learner.}}

\begin{tabular}{@{}lll@{}}
\toprule
Model             & Hyperparameter              & Value   \\ \midrule
Tappability Head  & Learning Rate               & 0.0005  \\
                  & Batch Size                  & 32      \\
                  & Hidden Size                 & 64      \\
                  & Num Layers                  & 4       \\
Draggability Head & Learning Rate               & 0.00005 \\
                  & Batch Size                  & 32      \\
                  & Hidden Size                 & 64      \\
                  & Num Layers (Self-Attention) & 1       \\
                  & Num Layers (Classifier)     & 1       \\
Screen Similarity & Learning Rate               & 0.00001 \\
                  & Batch Size                  & 64      \\ \bottomrule
\end{tabular}
\label{tab:hyperparams}
\end{table*}

\end{document}